\documentstyle[times,graphics,epsfig,amsmath,amssymb,natbib,psfrag,color]{mn2e}

\newcommand{\be}{\begin{equation}}
\newcommand{\e}{\end{equation}}
\newcommand{\bear}{\begin{eqnarray}}
\newcommand{\ear}{\end{eqnarray}}

\def\xh1{x_{H~{\sc i}\,}\,}
\def\apj{ApJ}

\def\apl{ApJL}
\def\mnras{MNRAS}

\def\nat{Nature}
\def\u{{\vec U}}
\def\th{\vec{\theta}}

\def\HI{H~{\sc i}\,}
\def\HII{H~{\sc ii}\,}
\begin{document}
   \title[Detecting anisotropic ionized bubbles]{The impact of 
   anisotropy from finite light travel time on detecting ionized bubbles
   in redshifted 21-cm maps.}

\author[Majumdar, Bharadwaj, Datta \& Choudhury  ]
{Suman Majumdar$^1$\thanks{E-mail: sumanm@phy.iitkgp.ernet.in},
Somnath Bharadwaj$^1$\thanks{E-mail: somnath@phy.iitkgp.ernet.in},
Kanan K. Datta$^2$\thanks{E-mail: kdatt@astro.su.se }
and T. Roy Choudhury$^3$\thanks{E-mail: tirth@hri.res.in}\\
$^1$Department of Physics and Meteorology \& Centre for Theoretical Studies,
 IIT, Kharagpur 721302, India\\
$^2$The Oskar Klein Centre for Cosmoparticle Physics\& Department of Astronomy,
 Stockholm University, Albanova, SE-10691 Stockholm, Sweden\\
$^3$Harish-Chandra Research Institute, Chhatnag Road, Jhusi, Allahabad 211019,
 India}
\maketitle

\begin{abstract}
The detection of ionized bubbles around quasars in redshifted 21-cm maps
is possibly one of the most direct future probes of reionization. 
 We consider two   models for the growth of   spherical ionized
 bubbles  to study the   apparent shapes   of the bubbles in
 redshifted 21-cm maps,  taking into  account the finite light travel
 time (FLTT) across the   bubble. In both  models the bubble has a
 period of rapid growth beyond which its  radius either 
saturates or
 grows slowly. We find that the  FLTT, whose   effect is particularly
 pronounced for large bubbles,  causes the   bubble's image  to
 continue to  grow  well after  its actual growth is over.   
There are
 two distinct FLTT distortions in the bubble's  image: (i)  its 
 apparent center is  shifted along the line of  sight (LOS)  towards
 the observer from the  quasar;   (ii) it is anisotropic along
 the  LOS.  The bubble  initially  appears  elongated  along the
 LOS. This  is reversed in the later stages of  growth 
 where  the bubble appears compressed.  

The FLTT distortions are  expected to have an impact on matched filter
 bubble detection where  it is most convenient to use a spherical
template for the filter. We find that the best matched spherical
 filter gives a reasonably good estimate of the size and the shift in
 the center of the anisotropic image. The mismatch between the
 spherical filter and the anisotropic image causes  a degradation in
 the  SNR relative to that of a spherical bubble. The degradation 
 is in the range   $10 -20 \%$ during the period of rapid growth when
 the  image appears elongated, and is less than $10 \%$ in the later
 stages  when the image appears compressed. 

We conclude that a spherical filter is adequate for bubble
detection. The FLTT distortions do not affect the lower limits for
bubble detection with $1000 \, {\rm   hr}$ of GMRT  observations. The
smallest spherical filter for which  a detection is possible has 
 comoving radii  $24 \, {\rm  Mpc}$ and $33 \, {\rm   Mpc}$ for a $3
 \sigma$ and $5 \sigma$ detection respectively, assuming a neutral
 fraction $0.6$ at $z \sim 8$.
\end{abstract}


\section{Introduction}
According to our current understanding, reionization of neutral hydrogen (\HI)
is most likely an extended process operating over redshifts 
$6 \lesssim z \lesssim 15$
and is most likely driven by stellar sources forming within galaxies
(for reviews, see \citealt{choudhury06,choudhury09}). The
\HI distribution at these epochs is characterized by ionized 
bubbles centered around galaxies,  with characteristic sizes
determined by the efficiency of photon production and the clustering
properties of the galaxies \citep{furlanetto1}.  These bubbles 
overlap as  reionization proceeds, and  the  \HII  (ionized hydrogen) 
 distribution develops a complex topology. Statistical probes of the
 redshifted   21-cm signal, like the  power spectrum (see
\citealt{furlanetto2} for a review) and the bispectrum
\citep{Bharadwaj05},  carry  signatures of these bubbles. 
A detection of  the statistical signal  would indirectly constrain the
bubble   distribution.  In contrast, it would
be  possible to directly probe this     by detecting the
individual \HII  regions.  

At this moment  it is not clear whether the first generation of 21-cm
observations would be able to detect  the detailed topology of \HII
regions and constrain 
reionization histories. Most likely, the present day instruments 
({GMRT\footnote{http://www.gmrt.ncra.tifr.res.in}} 
\citep{swarup} ) 
or those expected in the
near future (e.g., {MWA\footnote{http://www.haystack.mit.edu/ast/arrays/mwa/}},
 LOFAR\footnote{http://www.lofar.org/}) 
might just be adequate for detecting the
presence or absence of an ionized region. The first target could possibly be 
detection of \HII bubbles around targeted luminous sources
(say, quasars) which can have much larger sizes than the typical 
galaxy  generated ionized regions.
The growth of such a region may,  in the  simplest
 situation, be modeled  as  a spherical ionized bubble  embedded in a  
 uniform neutral medium and its growth can be followed
 analytically \citep{shapiro87}.  
The redshifted \HI 21-cm signal from such a  bubble will be buried in
 foregrounds and noise, both of which are considerably larger than the
 signal. 
Further the noise in different pixels of a radio interferometric image
 is correlated.  It is a big challenge to detect this faint redshifted 
21-cm  signal of an ionized bubble. 
\citet{datta2} (hitherto Paper I) have developed a visibility based
 matched filter 
 technique that  optimally combines the \HI signal of an extended bubble
 while removing foregrounds and minimizing the noise.  
The analysis shows that density fluctuations in the neutral hydrogen  
outside the bubble 
impose  a limit  on the comoving radius of the
 smallest bubble that can be  detected.  This limit is independent of
 the observing  time. Simulations show (\citealt{datta3}; hitherto
 Paper II) that 
 bubbles of  comoving radius $\le 6\, {\rm Mpc}$ and $\le 12\, {\rm Mpc}$ 
cannot respectively be  detected using the  
GMRT and the MWA respectively, however large be the observing time.  
In addition, it is also possible to show that the
 redshift $z \sim 8$  is  
optimum for bubble detection (\citealt{datta09}; hitherto Paper III). 
At the redshift $z=8$, for a bubble located
 at the center of the field of view (FoV),   it will be possible to
 detect ($3 \sigma$) a   bubble of comoving radius $\ge 19\, {\rm Mpc}$ and 
$\ge 22\, {\rm Mpc}$ with $1000$ hrs of observation 
 using the GMRT and MWA respectively provided the gas outside the bubble
 is completely neutral. This prediction is somewhat modified if a
 significant fraction of the neutral gas outside the bubble is ionized
 by $z \sim 8$. In a situation where
 the  neutral hydrogen fraction outside the bubble is 
$\xh1\sim 0.6$ at $z=8$ \citep{choudhury06, choudhury09},  the
 comoving radius of the smallest  bubble that can be detected in
 $1000$ hrs of observation   is $\sim 24\, {\rm Mpc}$ and 
$\sim 28\, {\rm Mpc}$ for the GMRT and MWA respectively.

Most of the analysis of matched filter bubble detection (Papers I, II 
and III) assumes  the ionized region to be a spherical bubble. However,
a growing spherical bubble will appear anisotropic for a present day
observer  due to  the finite light travel time (FLTT) 
\citep{wyithe04,yu05,wyithe05,shapiro06,sethi08}, and due to the 
evolution of the global ionized  fraction \citep{geil2}. 
 These anisotropies, if detected, would provide important information about
 the quasar luminosity and the evolution of the global ionization
 fraction.  While this is an interesting possibility,
 it is unlikely that it will be feasible to discern such details using
 either  the present day instruments (GMRT)  or those expected
 in the near future.(eg. MWA) which are just adequate for
 detecting the presence or absence of an ionized bubble.
 Given this, it would be most appropriate to search for ionized
 regions using a spherical bubble as a template in   the  matched
 filter technique. The search can either be completely blind where we
 vary  
the four parameters $[\theta_x,\theta_y,z_b,R_b]$, or it can be
targeted along a  
known quasar where it is sufficient to vary $[z_b,R_b]$.  Here 
$\theta_x,\theta_y,z_b$ are respectively the two angular
coordinates and redshift of the bubble center, while $R_b$ is
the  bubble's comoving radius. In the matched filter analysis,
the signal to noise ratio is maximum  when the parameters of the
filter exactly match those of the bubble  actually present in
the data. In other words, a detection is achieved by varying the
parameters of the filter so that it is exactly matched to the signal
of the bubble that is actually present in the data. The apparent
anisotropies will, however, introduce a mismatch between the signal and
the filters.  This is potentially a serious issue for both a blind
search and targeted   bubble detection.   In this paper we study
the impact of this apparent anisotropy on our ability to detect
ionized bubbles in a targeted search around a known quasar using the
matched  filter technique. 

In reality apart from these anisotropies the
ionized regions around quasars would not be spherical because of various
other effects like inhomogeneities in the IGM, overlapping ionized bubbles
arising from stellar sources and possible anisotropic emission from
the quasar itself.  
Moreover, the foreground subtraction could also subtract a part of the
signal,  which may change the recovered bubble shape. 
The anisotropy in bubble shape caused by these effects
would vary from bubble to bubble, and it is rather difficult to
incorporate these effects in our analytical estimations, so we do not
attempt this here.   \citet{geil1} and \citet{geil2} have studied
several of these effects using semi-numerical simulations. 

A brief outline of the paper follows. In Section 2. we discuss the
equations governing the growth of a spherical ionized bubble and the
apparent anisotropy due to the FLTT. Section 3. reviews the matched
filter technique for bubble detection.  In Section 4. we quantify the
apparent anisotropy due to the FLTT and assess its impact on bubble
detection. We also   discuss the effect of an evolving global neutral
fraction. We summarize and conclude in section 5.  Growth equation for
ionized bubble with an evolving neutral fraction has been discussed
in Appendix A. 

Unless mentioned otherwise, throughout the paper we present results 
for the redshift $z=8$ assuming the neutral hydrogen fraction 
$\xh1=0.6$ outside the bubble, with the cosmological parameters 
 $h= 0.74 $, $\Omega_m = 0.3$, $\Omega_{\Lambda} = 0.7$, $\Omega_b h^2
= 0.0223$. 

\section{Growth of Ionized Bubbles and Their Apparent Shape}
\label{sec:shape}
\begin{figure}
\psfrag{O}[c][c][1][0]{{\bf {\Large O}}}
\psfrag{Q}[c][c][1][0]{{\bf {\large Q}}}
\psfrag{A}[c][c][1][0]{{\bf {\large A}}}
\psfrag{t}[c][c][1][0]{{\bf {\Large $\phi$}}}
\includegraphics[width=.45\textwidth, angle=0]{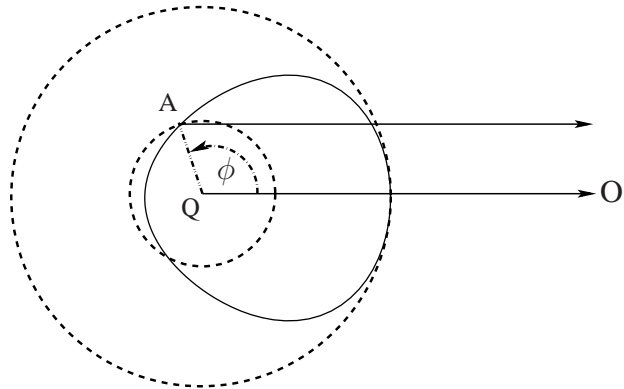}
\caption{The dashed lines show different stages of the growing \HII
  bubble around a quasar Q. The solid curve shows the apparent shape 
of the \HII bubble for a distant observer O located in  the direction
  indicated   by the arrows. }
\label{fig:shape}
\end{figure}

We first  review the equation governing the growth of a spherical
 \HII  region around a quasar  
\citep{shapiro87,white03,wyithe04,wyithe05,yu05,shapiro06,sethi08}.
  For our purpose we consider the form given in   eq. (7) 
of \citealt{yu05}, which is   
\begin{equation}
\frac{4\pi}{3} \frac{d}{d\tau}\left(\xh1 \langle n_H \rangle
r^3\right)=\dot{N}_{phs}(\tau)-\frac{4}{3}\pi \alpha_B C \langle n_H
\rangle ^2
r^3\,. 
\label{eq:yu}
\end{equation}
Here the quasar is assumed to be triggered at a cosmic time $\,\,t_i$,
and $\tau  = t - t_i$ denotes the quasar's age at any
later time $t$. The variable $r(\tau)$ denotes the radius of the spherical
ionizing front (Figure \ref{fig:shape}) at the instant when a
photon that was emitted from the quasar at $\tau$ catches up with the 
ionizing front. 
The quasar is assumed to emit  ionizing photons isotropically at a rate
$\dot{N}_{phs}(\tau)$ which  completely ionizes the hydrogen inside the
spherical  bubble of  radius $r(\tau)$. The bubble is  
surrounded by gas with  mean  hydrogen number density  $\langle n_H
\rangle$ and neutral fraction $\xh1$.  The term
$\alpha_B(=2.6\times 10^{-13} \, {\rm  cm}^3 \, s^{-1})$ is the 
recombination coefficient to excited levels of hydrogen  at  
$T=10^4  \, {\rm K}$, and  
$C \equiv \langle n^2_{\HI} \rangle/\langle n_H \rangle^2$ is the
clumping   factor, which quantifies the  effective clumpiness of the
hydrogen inside the bubble. Eq. (\ref{eq:yu}) essentially tells us
that the growth of the bubble is driven by the supply  of ionizing
photons after accounting for the photons required to compensate for
the recombinations inside the existing ionized region.  The effect of
Hubble expansion is not included  in this equation. 

Assuming that all the quantities except $r(\tau)$ and
$\dot{N}_{phs}(\tau)$ are constant over the time-scale of the bubble's
growth, we have the solution 
\begin{equation} 
r(\tau)=\left[\frac{3}{4 \pi \langle n_H \rangle
\xh1}\int\limits_0^{\tau}{\dot{N}_{phs}(\acute{\tau})\, 
\exp\left(\frac{\acute{\tau}-\tau}{\tau_{rec}}\right)\,
d\acute{\tau}}\right]^{\frac{1}{3}}     
\label{eq:sol}
\end{equation}  
where $\tau_{rec}$ is  the recombination time defined as
\begin{eqnarray}
\tau_{rec}&=& \xh1 \left(C\,\langle n_H
\rangle\,\alpha_B\right)^{-1}\nonumber\\ 
&\simeq&4 \times 10^6 \,{\rm yr}\left(\frac{\xh1}{0.1}\right) 
\left(\frac{30}{C}\right)\left(\frac{7.4}{1+z}\right)^3  \,.
\label{eq:taurec}
\end{eqnarray}
For the redshift of our interest ($z=8$)  and assuming that 
 $\xh1=0.6$ and $C = 30$ \citep{yu05b} we have 
 $\tau_{rec} \simeq 1.33 \times 10^7 {\rm yr}$ which we use
 throughout. 

Following \citealt{yu05},  we consider two different models for  the photon
emission rate $\dot{N}_{phs}(\tau)$. 
In Model $(i)\, \, \dot{N}_{phs}(\tau)=\dot{N}_{phs,i}$ is a constant
whereas it increases exponentially in 
 Model  $(ii)\, \, \dot{N}_{phs}(\tau)\,= \dot{N}_{phs,i} \,
 \exp\left(\tau/\tau_S \right)$.  Here    
 $\tau_S\simeq4.5\times10^7 \, {\rm   yr} \, 
 \left[\frac{\epsilon}{0.1(1-\epsilon)}\right]\,$ where   $\tau_S$ is
 the  Salpeter time-Scale and $\epsilon$ is the mass to energy
 conversion  efficiency whose value  typically is    $\epsilon \simeq 0.1$ 
which we adopt throughout \citep{ yu02, yu04, yu05b}. This gives 
$\tau_s \simeq 5 \times 10^7 \, {\rm yr}$ which we have
used throughout in Model $(ii)$. 
\begin{figure}
\psfrag{r}[c][c][1][0]{{\LARGE $r(\tau)/r_s$}}
\psfrag{t}[c][c][1][0]{{\LARGE $\tau /\tau_{rec}$}}
\includegraphics[width=.35\textwidth, angle=-90]{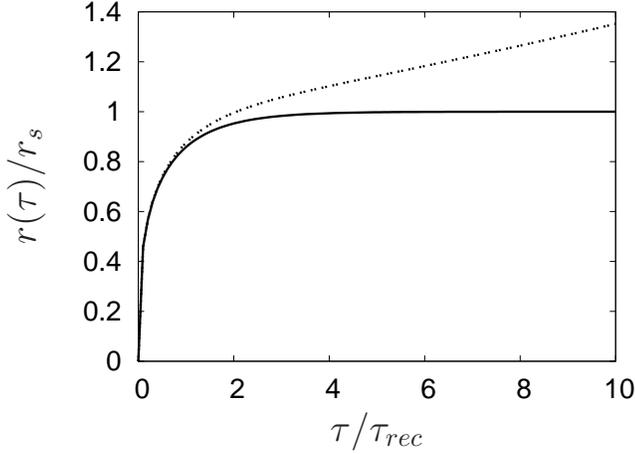}
\caption{The  growth of an \HII bubble for Models (i) and
  (ii) shown  by the  solid and dashed curves respectively.}
\label{fig:r_tau}
\end{figure}
The bubble growth  $r(\tau)$  for model $(i)$ and $(ii)$ are
respectively given by
\begin{equation}
r(\tau)=r_S \left[1-\exp\left(-\frac{\tau}{\tau_{rec}}\right)
\right]^{\frac{1}{3}}   
\label{eq:sol1}
\end{equation}
and
\begin{equation}
r(\tau)=\,r_S \, 
\left\{\left(\frac{\tau_s}{\tau_s\,+\,\tau_{rec}}\right)
 \left[\exp\left(\frac{\tau}{\tau_s}\right)-
\exp\left(-\frac{\tau}{\tau_{rec}}\right)\right] \right\}^{\frac{1}{3}} \,    
\label{eq:sol2}
\end{equation}
where
\begin{equation}
r_S=\left(\frac{3\dot{N}_{phs,i}\,\tau_{rec}}{4\,\pi\,\xh1\,
  \langle n_H \rangle} \right)^{\frac{1}{3}} 
\label{eq:rs}
\end{equation}
for which the solutions are shown in Figure \ref{fig:r_tau}.
In both models the bubble  has an initial  period 
$(\tau < \tau_{rec})$ 
of rapid  growth $r(\tau)\simeq (\tau/\tau_{rec})^{1/3}$. 
 In Model $(i)$ the bubble radius subsequently 
approaches a constant value $r(\tau)=r_s$. In Model $(ii)$ the bubble
radius does not reach a steady value but continues to grow
at a  slower rate beyond $r > r_s$. 

To visualize the apparent shape of the bubble as seen 
 by a  present day observer we  need
the relation between $r$ and $\phi$, where $\phi$
 is the angle between observer's line of sight (LOS) and
 the point A under consideration on the ionization front 
 (Figure \ref{fig:shape}). The light travel time starting from
 the quasar at $\tau$ to the point A and then to the present day
 observer is $[r(\tau)/c](1-\cos \phi)$ more  compared  to the
 photon that was emitted from the quasar at age $\tau_Q$ and travels
 straight to the present day observer. This gives 
\begin{equation}
\tau_Q=\tau +\frac{r(\tau)}{c}(1-\cos\,\phi) \,.
\label{eq:shape}
\end{equation} 
(eq. (3) of \citealt{yu05}) 
which we use   along with eq. (\ref{eq:sol1}) or eq.
(\ref{eq:sol2}) to determine $r$ as a function of $\phi$ for Models
$(i)$ and $(ii)$ respectively. 
\begin{figure}
\psfrag{rmpc}[c][c][1][0]{{\bf {\Large $r\,$ Mpc}}}
\psfrag{theta}[c][c][1][0]{{\bf {\Large $\phi \,$ degree} }}
\includegraphics[width=.17\textwidth,angle=270]{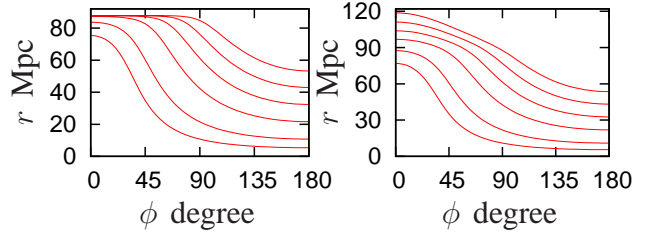}
\caption{The variation of $r$  with $\phi$ for  Models
 $(i)$  (left) and $(ii)$ (right))  with $r_s=88\,{\rm
 Mpc}$.   In both  panels 
 $\tau_Q$  takes values $ (1.3,2.7,5.3,8.0,10.6,13.3)\times 10^7 \,
 {\rm yr}$  from the bottom to top. The values of  $r_s$ and $r$ are both
in comoving Mpc.}  
\label{fig:r_theta}
\end{figure}
\begin{figure}
\psfrag{Mpc}[c][c][1][0]{{\bf {\Large Mpc}}}
\includegraphics[width=.21\textwidth,angle=270]{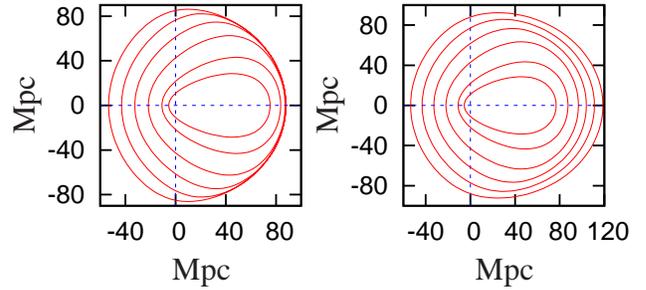}
\caption{The shape of the \HII bubble as it would appear in
  a redshifted 21-cm map. The distant  observer is located to the
  right of the   figure, and   the line of sight is along the $x$
  axis.  The quasar   is located   at the intersection of the two
  dashed lines.   The   photon emission model, $r_s$ and $\tau_Q$ values 
  are the same as in Figure \ref{fig:r_theta}. The left and right panels
correspond to Model $(i)$ and $(ii)$ respectively.} 
\label{fig:shape_all} 
\end{figure}

The photons from  the  front part   of the bubble ( $\phi < 90^{\circ}$) take
 less time to  reach the observer as compared to photons from the
 rear part  ($\phi > 90^{\circ}$).    As a consequence the  observer sees
 different  parts of the bubble at different stages of
 its growth whereby it appears anisotropic along the LOS.

The apparent shape of the  \HII region is controlled by the parameters
$r_s\,$ and $\tau_Q$, where $\tau_Q$ is the quasar's age as seen by a
present day observer.  In Figure \ref{fig:r_theta} we have explicitly 
shown $r$ as a function of $\phi$ for $r_s = 
8 \,\times\, c \tau_{rec} (1+z) \ =88 \, {\rm Mpc}$ (comoving) 
and $\tau_Q/\tau_{rec}=1, 2, 4, 6, 8, 10 $, which correspond to
$\tau_Q=(1.3, 2.7,5.3,8.0,10.6,13.3)\times 10^7\, {\rm yr}$
respectively. In Figure  \ref{fig:shape_all} we show  a
radial section through the center of the  bubble as it would appear in 
a redshifted 21-cm map.  The difference in light travel time across
the bubble is more for larger bubbles, which is why we have chosen a
particularly large value of $r_s$  to illustrate the anisotropy. 
Note that here and in all subsequent discussion  we
use comoving length-scales.

The bubble  grows  rapidly  when $\tau_Q/\tau_{rec} \le 1$ which 
 is  close to the instant when the quasar was  triggered,  We then
 have a  large difference between the front and back 
surfaces (Figure \ref{fig:r_theta}). The back surface, is viewed at
an earlier phase of growth compared to the  front
surface. This has mainly two  effects on the bubble's apparent shape 
(Figure  \ref{fig:shape_all}), (i) the  center is shifted along the
 LOS towards the observer, (ii) the bubble  appears anisotropic.  We
 also note that the bubble appears elongated along the 
 LOS in the early stage of growth. The difference between the front
 and back surfaces gradually comes down with increasing $\tau_Q$. The
 apparent center then approaches the quasar position, and the
 elongation along the LOS also diminishes. While Models $(i)$ and $(ii)$
 are nearly indistinguishable at small  $\tau_Q$,  the behavior are
 somewhat different at  large $\tau_Q$.

\begin{figure}
\psfrag{LOS}[c][c][1][0]{{\bf {\large LOS}}}
\psfrag{Q}[c][c][1][0]{{\bf {\large Q}}}
\psfrag{A}[c][c][1][0]{{\bf {\large A}}}
\psfrag{B}[c][c][1][0]{{\bf {\large B}}}
\psfrag{C}[c][c][1][0]{{\bf {\large C}}}
\psfrag{t}[c][c][1][0]{{\bf {\Large $\phi_c$}}}
\includegraphics[width=.45\textwidth, angle=0]{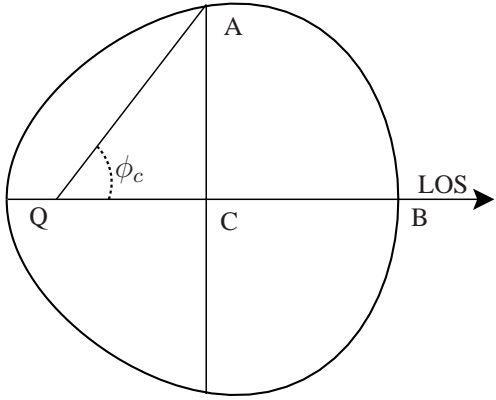}
\caption{This shows the apparent shape of an \HII bubble as it would
  to a distant observer located along the line of sight (LOS)
  indicated in the figure. The apparent center of the bubble shifts
by  $\overline{QC} = \Delta r/2$  from the quasar Q to  the point
  C. The apparent radius is   $\overline{CB} = R_{\parallel}$ along
  the LOS and $\overline{CA} = R_{\perp}$ in the perpendicular direction.}
\label{fig:eta_dia}
\end{figure}

\begin{figure}
\psfrag{tq}[c][c][1][0]{{\LARGE $\tau_Q/10^7$ }}
\psfrag{rs}[c][c][1][0]{{\LARGE $r_s$ Mpc}}
\psfrag{20}[c][c][1][0]{{\Large$20$}}
\psfrag{30}[c][c][1][0]{{\Large$30$}}
\psfrag{40}[c][c][1][0]{{\Large$40$}}
\psfrag{50}[c][c][1][0]{{\Large$50$}}
\psfrag{60}[c][c][1][0]{{\Large$60$}}
\psfrag{70}[c][c][1][0]{{\Large$70$}}
\psfrag{80}[c][c][1][0]{{\Large$80$}}
\psfrag{90}[c][c][1][0]{{\Large$90$}}
\psfrag{100}[c][c][1][0]{{\Large$100$}}
\includegraphics[width=.8\textwidth,angle=270]{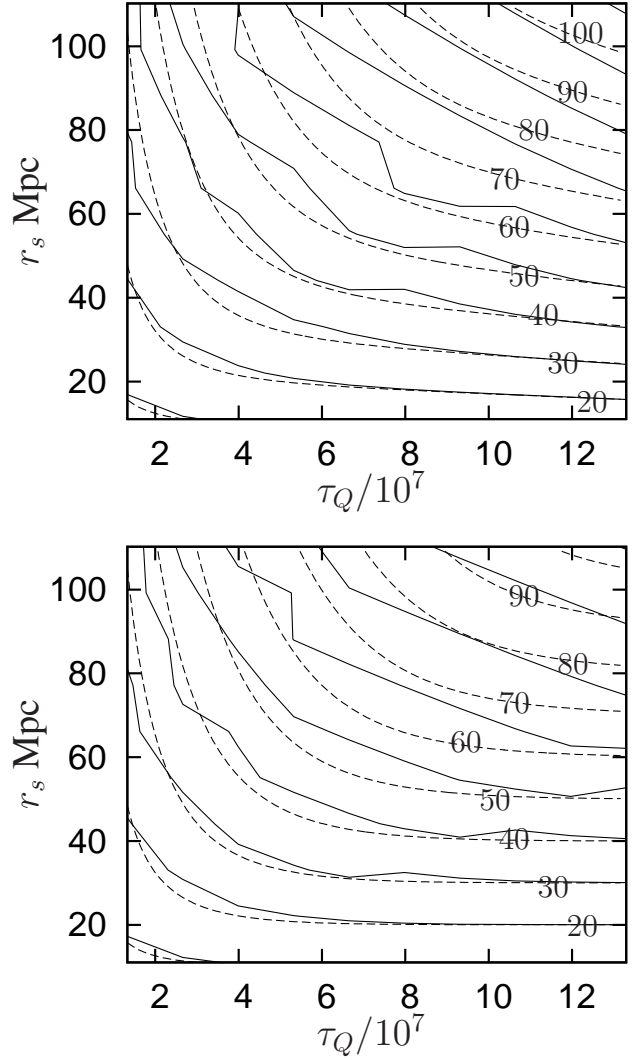}
\caption{Contours of  $R_{\perp}$ (dashed lines) 
and $R_f$ (solid lines) in comoving ${\rm Mpc}$, shown 
for Models $(i)$ and $(ii)$ in the bottom and top panels respectively.
$R_f$ (defined in Section 4) is the radius of the spherical bubble
that best matches an  anisotropic bubble with parameters  $r_s$ and
$\tau_Q$.} 
\label{fig:r_perp}
\end{figure}

\begin{figure}
\psfrag{tq}[c][c][1][0]{{\LARGE $\tau_Q/10^7$ yr}}
\psfrag{rs}[c][c][1][0]{{\LARGE $r_s$ Mpc}}
\psfrag{0.2}[c][c][1][0]{{\Large$0.2$}}
\psfrag{0.4}[c][c][1][0]{{\Large$0.4$}}
\psfrag{0.6}[c][c][1][0]{{\Large$0.6$}}
\psfrag{0.8}[c][c][1][0]{{\Large$0.8$}}
\includegraphics[width=.8\textwidth,angle=270]{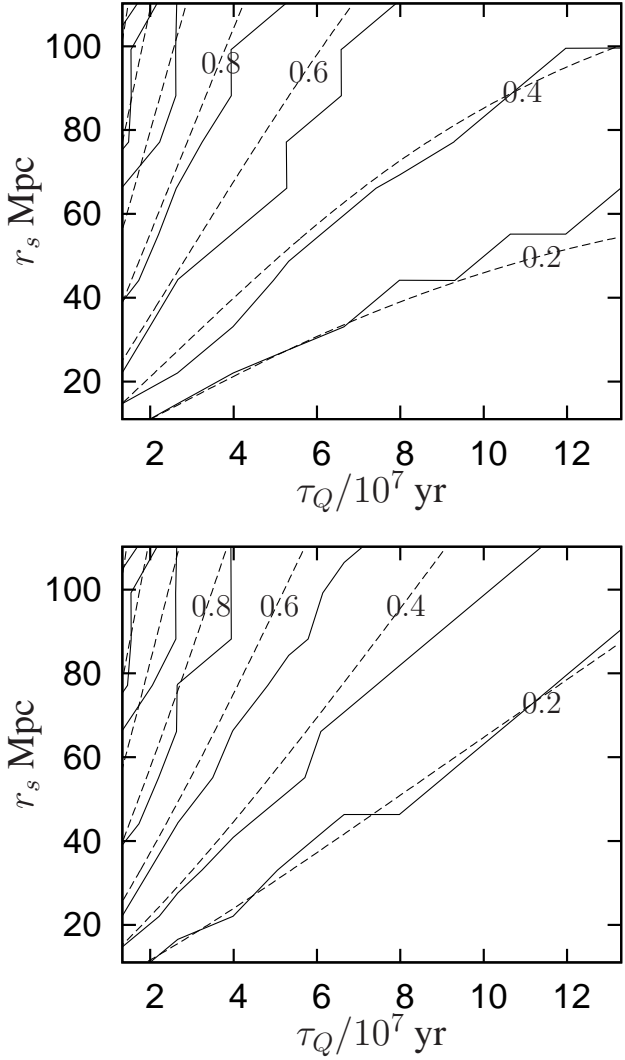}
\caption{The dashed lines show contours of the shift parameter 
$s$ defined by Eq. (\ref{eq:s}).  The solid lines show contours of 
$s_o$  which quantifies  the shift  between the
  center of the best   matched   filter and the quasar (defined in
  Section 4.).  Models $(i)$ and $(ii)$ are shown
  in the   bottom and top panels respectively.} 
\label{fig:shift}
\end{figure}

\begin{figure}
\psfrag{tq}[c][c][1][0]{{\LARGE $\tau_Q/10^7$ yr}}
\psfrag{rs}[c][c][1][0]{{\LARGE $r_s$ Mpc}}
\psfrag{-0.2}[c][c][1][0]{{\Large$-0.2$}}
\psfrag{-0.1}[c][c][1][0]{{\Large$-0.1$}}
\psfrag{0.0}[c][c][1][0]{{\Large$0.0$}}
\psfrag{0.1}[c][c][1][0]{{\Large$0.1$}}
\includegraphics[width=.8\textwidth,angle=270]{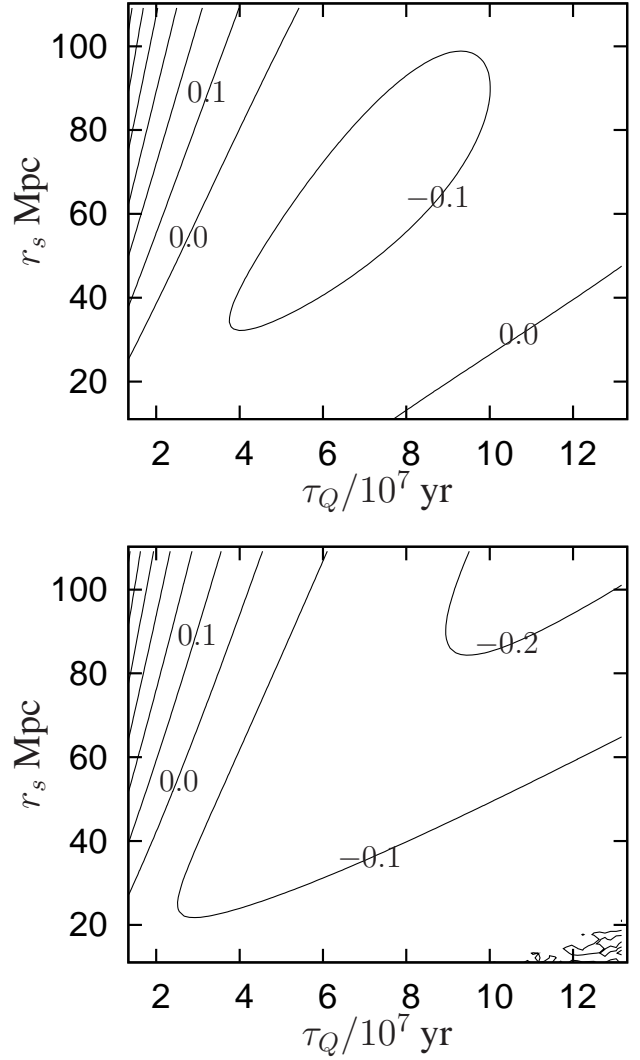}
\caption{This shows  contours of the anisotropy parameter $\eta$
 (defined by Eq. (\ref{eq:eta})) 
 for Models $(i)$ and $(ii)$ in the bottom and top panels respectively.
 The contours are at an interval of $0.1$. }
\label{fig:eta}
\end{figure}

For a fixed  $r_s$ and $\tau_Q$, we 
 use $r_{min}$ and $r_{max}$ to denote the bubble's comoving  radius
 $r$ at  $\phi=180^{\circ}$ and $0^{\circ}$ respectively.  
The difference $\Delta r=r_{max}-r_{min}$ causes  the
bubble's center to shift by a comoving distance $\Delta r/2$ relative
 to the  quasar along the LOS  towards the observer (Figure 
\ref{fig:eta_dia}). The  bubble has  comoving radii
$R_{\parallel}=(r_{max}+r_{min})/2$  and $R_{\perp}$ 
along the LOS and perpendicular to it  respectively
(Figure \ref{fig:eta_dia}). In the subsequent discussion we use
 $R_{\perp}$ to characterize the comoving size of the anisotropic
 bubble. The dimensionless ratio 
\begin{equation}
s=\frac{\Delta r}{2 R_{\perp}} 
\label{eq:s} 
\end{equation}
expresses the shift in the bubble's center as a
 fraction of its radius $R_{\perp}$.  Further, we use
\begin{equation} 
 \eta=\frac{R_{\parallel}}{R_{\perp}}-1
\label{eq:eta}
\end{equation}
  to quantify   the anisotropy of
 the bubble. A value $\eta >0$ indicates that the bubble appears
 elongated along the LOS, whereas $\eta <0$ indicates that it appears 
compressed along the LOS. We have calculated $R_{\perp}$, $s$ and
 $\eta$ for a 
 range of $r_s$ and $\tau_Q$, for which the results are shown in the
 contour plots of Figures \ref{fig:r_perp},  \ref{fig:shift}  and 
 \ref{fig:eta} respectively.   The $r_s$, $\tau_Q$ range was chosen
 so that  the largest   bubble has a comoving radius roughly in the
 range  $70-80 \,  {\rm Mpc}$. 


In our models the bubble, as viewed in the quasar's rest-frame, has a
short period of rapid growth $\tau < 10^7 \, {\rm yr}$ beyond which it
saturates (Model (i)) or grows slowly (Model (ii)) (Figure \ref{fig:r_tau}).
Naively one would expect the bubble's image in redshifted 21-cm maps
to exhibit a similar behavior. However, we see that the bubble's
image continues to grow (Figure \ref{fig:r_perp}) well beyond  $\tau =
10^7 \, {\rm yr}$. The effect is more for larger bubbles 
where  the light travel time is longer. The shift and anisotropy also
are non-zero well beyond the period of the bubble's actual growth. For
a fixed $r_s$, the shift decreases monotonically with increasing
$\tau_Q$. While the shift can be as large as $s \ge 1$ (Figure
\ref{fig:shift}) for  a large   bubble  seen in an early  phase  of its
growth, the  typically value of $s$ is  in the range $0.2 \le s \le 
0.6$.  The bubbles appear elongated ( $\eta >0$). in the early phase of
their  growth (Figure \ref{fig:eta}). For a fixed $\tau_Q$, we have
more anisotropy  for a larger bubble. The elongation diminishes with
increasing $\tau_Q$. This is followed by a situation where the bubble
appears  compressed $(\eta < 0)$ along the LOS. This occurs 
when  a bubble is viewed in the late stage of its  growth  where  most
of the bubble, except a small part of the back surface, has nearly
stopped growing.    The transition from an  elongated bubble to a
compressed bubbles is clearly visible  in Figure \ref{fig:shape_all}.  
For  both Models $(i)$ and $(ii)$, by and large,  we expect  the bubbles
to be compressed  with anisotropies of the order of $\mid \eta \mid
\approx 0.1$ in Model $(i)$, and somewhat larger anisotropies in Model
$(ii)$. Elongated bubbles may have anisotropies of the order
$\eta \approx 0.1 - 0.5$, but these will be seen  only in bubbles
surrounding recently triggered quasars.

\section{Matched Filter Bubble Detection Technique}
The quantity measured in radio-interferometric 
observations is the visibility $V(\u,\nu)$ which is
 related to the specific intensity pattern on
 the sky $I_{\nu}(\th)$ as 
\begin{equation}
V(\u,\nu)=\int d^2 \phi A(\th) I_{\nu}(\th)
e^{ 2\pi \imath \th \cdot \u}
\label{eq:vis}
\end{equation}
Here the baseline $\u={\vec d}/\lambda$ denotes 
the antenna separation ${\vec d}$  projected in the plane 
perpendicular to the line of sight  in units of the observing
wavelength $\lambda$, $\th$ is a two dimensional vector in the plane
of the sky with origin at the center of the FoV
 ({\it i.e.} phase center), and $A(\th)$ is the 
beam  pattern of the individual antenna. 
For the GMRT this can be well 
approximated by Gaussian 
$A(\th)=e^{-{\theta}^2/{\theta_0}^2}$ 
where 
$\theta_0 \approx 0.6 ~\theta_{\rm FWHM}$.

The visibility recorded in a radio-interferometric
observations is actually a combination of several contributions 
\begin{equation}
V(\vec{U},\nu)=S(\vec{U},\nu)+HF(\vec{U},\nu)+N(\vec{U},\nu)+F(\vec{U},\nu)
\,. 
\label{eq:vis_comp}
\end{equation}
where the first term $ S(\u, \,\nu)$ is the expected signal of the
ionized region that we are trying to detect. 
Note that the signal from an ionized region appears as a decrement
with respect to the uniform background 21-cm radiation. 
This uniform background is however recorded only by the  baseline at zero 
spacing which is usually not considered in radio-interferometric
observations. 
Readers are referred to
Paper I (section 2) for  more detail regarding how to calculate  $
S(\u, \,\nu)$.   
$HF(\vec{U},\nu)$ is the contribution from  fluctuations in the 
 \HI distribution  outside the ionized bubble,  
$N(\vec{U},\nu)$ and $F(\vec{U},\nu)$ are the noise and
 foreground contributions respectively.

The signal  from an ionized bubble  is expected to be
 buried deep under the other contributions (noise and foreground)
which typically are orders of magnitude larger. We introduce an
 estimator $\hat{E}[\theta_x,\theta_y,z_b,R_b]$ 
 to search if a  particular  ionized bubble (with bubble parameters
$[\theta_x,\theta_y,z_b,R_b]$ ) is  present in our  observation.   
The estimator combines  all  the visibilities weighted by a filter 
$S_{f}(\u,\nu)[\theta_x,\theta_y,z_b,R_b]$, 
\begin{equation}
\hat{E}= \sum_{a,b} S_{f}^{\ast}(\u_a,\nu_b)
\hat{V}(\u_a,\nu_b) \, .
\label{eq:estim0}
\end{equation}
Note that in order to keep the notation compact we do not explicitly
show the parameters $[\theta_x,\theta_y,z_b,R_b]$ and
$\hat{V}(\u_a,\nu_b)$ represents the observed visibility.
The filter  is  designed (Paper I) to optimally combine the signal
corresponding the the bubble that we are trying to detect while
minimizing the contribution from the other  contaminants. The
expectation value of $\hat{E}$ is 
\begin{equation}
\langle \hat{E} \rangle=\sum_{a,b} S_{f}^{\ast}(\u_a,\nu_b) \,
S(\u_a,\nu_b)\, .  
\end{equation}
The other contributions to $\hat{V}$ are assumed to be random
variables of zero mean, uncorrelated to the filter, and hence they
contribute only to the variance  
\begin{equation}
\langle (\Delta \hat{E})^2 \rangle = \langle (\Delta \hat{E})^2
\rangle_{HF} +\langle (\Delta \hat{E})^2\rangle_{N} 
+ \langle (\Delta \hat{E})^2\rangle_{F} 
\end{equation}
where the subscripts $HF,N,F$ respectively refer to the contributions
from the \HI fluctuations, noise and foregrounds. The signal to noise
ratio for the estimator is defined as 
\begin{equation}
{\rm SNR}=\frac{\langle \hat{E} \rangle}{\sqrt{\langle (\Delta
    \hat{E})^2 \rangle}}
\end{equation} 

Bubble detection will be carried out by analyzing the SNR for
filters with different values of the  parameters
$[\theta_x,\theta_y,z_b,R_b]$. If an ionized bubble is actually
present in the FoV, the SNR will peak when the filter
parameters exactly match the  parameters of  the bubble. We shall have
a statistically significant bubble detection if the peak ${\rm SNR}
\ge 3$ or ${\rm SNR} \ge 5\,$ for a $3 \sigma$ or $5 \sigma$ detection
respectively. 

While the filter assumes the bubble to be spherical, it is evident
from the previous section that  bubbles are expected to appear
anisotropic   in redshifted 21-cm maps. It is in principle
possible to introduce the anisotropy as an additional search
parameter. This, however, would be required for  preliminary 
bubble detection  only if there is a severe mismatch between the  
spherical filter and the anisotropic image resulting in considerable
degradation of the SNR.

\section{Results}
\begin{figure}
\psfrag{Mpc}[c][c][1][0]{{\large Mpc}}
\psfrag{Rf}[c][c][1][0]{{\large $R_f\,{\rm Mpc}$}}
\psfrag{nuf}[c][c][1][0]{{\large $\nu_f\,{\rm MHz}$}}
\includegraphics[width=.55\textwidth,angle=270]{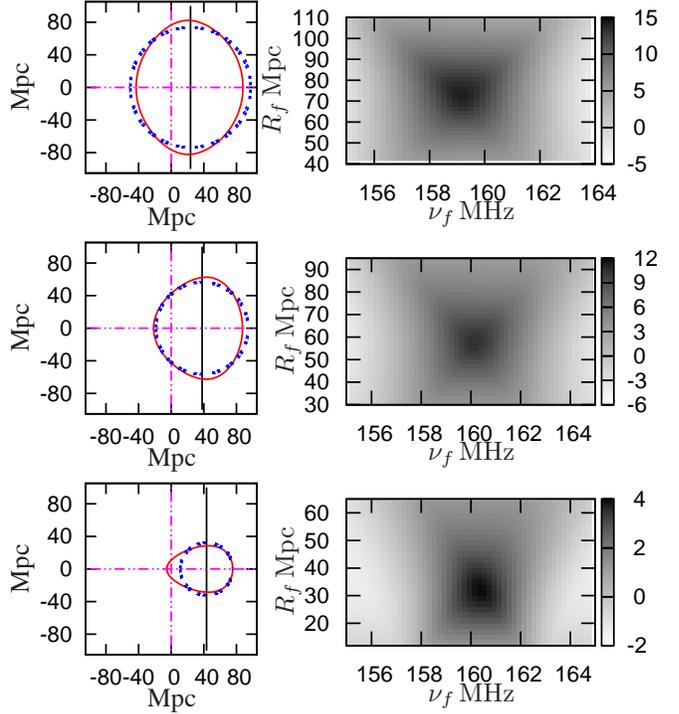}
\caption{This shows anisotropic shapes and corresponding expected SNR
 contours for Model $(i)$ considering 1000 hrs of observation with GMRT.
 From bottom to top solid curves in the left panels correspond to 
different apparent shapes of ionized regions for 
$\tau_Q = (1.3,\, 5.3,\, 10.6) \times 10^7\,{\rm yr}$ respectively with 
$r_s=88\, {\rm Mpc}$.
The dotted curves represent spherical filters for which
 the SNR is maximum for the shape in that corresponding panel.
 Intersecting point of the dash-dotted lines
 represent position of the quasar and intersecting point of the 
dash-dotted and solid lines represent position of the filter center. 
Right panels show the matched filter 
SNR contours for corresponding left panel anisotropic shapes as function 
of filter radius $R_f$ and filter center frequency coordinate $\nu_f$.}
\label{fig:snr_rs8_m1}
\end{figure}
\begin{figure}
\psfrag{tq}[c][c][1][0]{{\LARGE $\tau_Q/10^7$ yr}}
\psfrag{rs}[c][c][1][0]{{\LARGE $r_s$ Mpc}}
\psfrag{0.85}[c][c][1][0]{{\large$0.85$}}
\psfrag{0.90}[c][c][1][0]{{\large$0.90$}}
\psfrag{0.95}[c][c][1][0]{{\large$0.95$}}
\psfrag{1.0}[c][c][1][0]{{\large$1.0$}}
\psfrag{3s}[c][c][1][0]{{\Large$3 \sigma$}}
\psfrag{5s}[c][c][1][0]{{\Large$5 \sigma$}}

\includegraphics[width=.8\textwidth,angle=270]{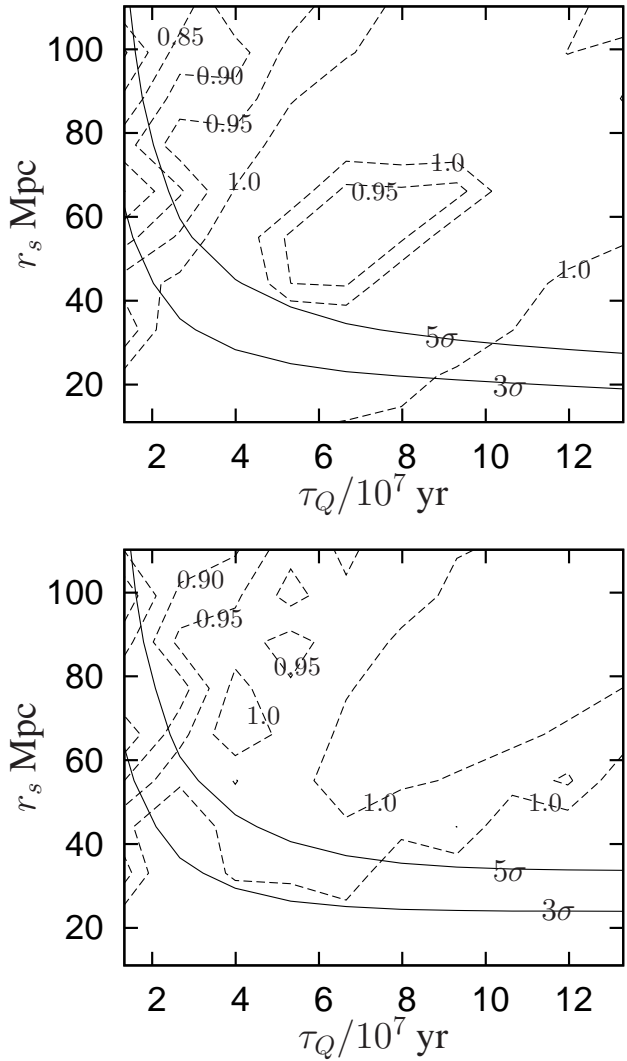}
\caption{The dashed  contours  show $f$ (defined in Section 4.), while
  the solid contours demarcate the regions where a $3 \sigma$ and 
$5 \sigma$  detection is possible. The  bottom and top panels
 show  results for Models $(i)$ and $(ii)$  respectively.} 
\label{fig:snr_frac}
\end{figure}
In order to keep the computational time of the
present investigation within bounds we consider a targeted search
along the LOS to a known high $z$ quasar.  The FLTT does not affect
the angular position of the bubble's  center. and  $\theta_x,\theta_y$
will coincide with the angular position of the quasar. We 
consider a search varying the parameters $R_f$  and $\nu_f$ which
respectively 
correspond to  the filter radius and the position of the filter's
center along the LOS. In the bottom panels of 
Figure \ref{fig:snr_rs8_m1} we have explicitly shown the the result
for a situation where a bubble with parameters $r_s=88 \, {\rm Mpc}$
and $\tau_Q=1.3 \times 10^7 \, {\rm yr}$ is present surrounding a
quasar which is located at $\nu_Q=158 \, {\rm MHz}$ in the center of 
the FoV.  The right panel shows the SNR for a  search in the
parameters 
$\nu_f$ and $R_f$. We find that the SNR peaks at $\nu_f=160.4 
\, {\rm MHz}$ and $R_f=32 \, {\rm Mpc}$, with ${\rm SNR }=3.99$ for
$1000 \, {\rm hr}$ of observation. The shape
of the best match filter is  shown in the left panel along with
the bubble's apparent shape. The middle  and top panels  consider the
same bubble at  two  later  stages of its growth.  We see that the
bubble, which appears elongated ($\eta=0.48$) in the bottom panel,
becomes compressed in the middle and top panels ($\eta=-0.11,-0.2$
respectively). In all cases the best matched filter provides a
reasonably good representation  of the bubble that  is actually
present. The mismatch  between the filter  and the bubble that is
actually present causes the SNR to be lower than that which is
expected if the image of the bubble were  spherical. 

We have considered bubble detection for a range of $r_s$ and $\tau_Q$
values for which  the radius of the best match filter $R_f$ are shown
in Figure \ref{fig:r_perp}. Figure \ref{fig:shift} shows 
the observed shift $s_o = (r_Q-r_c)/R_f$  where $r_Q$ and
$r_c$ are  the comoving distances to the quasar and the center of the
filter respectively.  We see that for most of  the parameter space 
$R_f$ and $s_o$  for  the best match filter closely follow $R_{\perp}$
and $s$ of the corresponding bubble image.   

The situation where both the  bubble and the corresponding quasar have
been  detected presents an interesting possibility. Both $R_f$ and
$s_o$ are known in such a situation. The curves corresponding to
constant $R_f$ and $s_o$ are non-degenerate, and it is possible  to
locate an unique point in the $r_s-\tau_Q$ parameter space once $R_f$
and $s_o$ are known. This holds the possibility of allowing us to
estimate the age of the quasar responsible for the bubble.  While the
age estimate would  depend  on the model assumed  for the
photon emission rate, our analysis considering two different models
indicates that we do not expect this variation to be very large.   

The mismatch between the anisotropic bubble and the  spherical filter 
is expected to cause a reduction in the SNR. For the best match filter 
Figure \ref{fig:snr_frac} shows  the ratio $f$ of the SNR for an 
anisotropic bubble to the SNR for a spherical bubble with the same
$R_f$.   We find that this ratio has value in the range $f \approx 0.9-1.0$
for nearly  the entire parameter space that we have considered. There
is a small region with small $\tau_Q$, which corresponds to rapid
bubble growth, where the ratio is in the range $f \approx 0.8-0.9$. We
note that 
this matches with  the region where the  anisotropy is  large  $\eta
\ge 0.4$. For
$1000\, {\rm hr}$ of observations with the GMRT, the $r_s$, $\tau_Q$
range for $3 \sigma$ and $5\sigma$ detection are shown in Figure
\ref{fig:snr_frac}.  The smallest filter radius (comoving) for  a
$3\sigma$ and a $5\sigma$ detection are  approximately  $24 \, {\rm
  Mpc}$ and $33 \, {\rm   Mpc}$ respectively, which   matches with the
corresponding values when the finite light travel  time is not taken
into account.
\subsection{Anisotropy due to evolving  $\xh1$}
\begin{figure}
\psfrag{z}[c][c][1][0]{{\LARGE $z$}}
\psfrag{xh1}[c][c][1][0]{{\LARGE $\xh1$}}
\includegraphics[width=.34\textwidth, angle=-90]{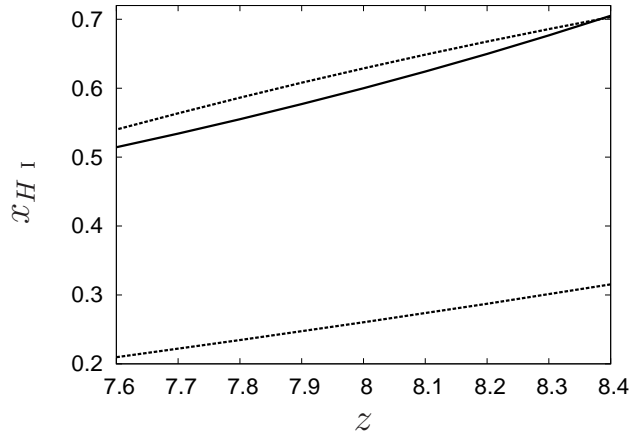}
\caption{This shows the evolution of $\xh1$ across the 
redshift range that is relevant for the growth and detection of a
bubble with $z_Q=8$. The lower and upper dotted curves respectively
show the  early and late reionization models  proposed by
\citet{choudhury05,   choudhury06}.  
The solid curve shows the model adopted for our analysis.}
\label{fig:xh1_z}
\end{figure}
The entire analysis, till now,  assumes a fixed  neutral fraction
$\xh1=0.6$ outside the bubble. The apparent anisotropy of a spherical
bubble will, in principle, be affected by the evolution of  $\xh1$.
The bubble's front surface which is  seen at a lower  redshift will
have a lower neutral fraction compared  to the back surface. This will
further increase the difference between the radius of the  front and
back surfaces  as  compared to the situation where $\xh1=0.6$ for
which  the results have been shown in Figure \ref{fig:r_theta} and
\ref{fig:shape_all}.  

The evolution of $\xh1$ during  reionization is largely unknown,  and
there exists  a wide range  of 
evolution histories all of which are consistent with the presently
 available observational constrains.  The lower and upper dotted
 curves in Figure  \ref{fig:xh1_z} respectively  shows the  early and
 late reionization models proposed by \citet{choudhury05,
   choudhury06}. The entire region bounded by these two curves is
 consistent with the current observational constraints. To assess the
 impact of $\xh1$ evolution on bubble detection we have considered a
 simple  model (the solid curve in Figure  \ref{fig:xh1_z}) where
 $\xh1^{-1}$ evolves linearly in the vicinity of $z=8$.  We do not
 expect $\xh1$ to decline very rapidly around $z=8$, and the slope has
 been chosen so as to approximately match those of the two dotted
 curves in Figure  \ref{fig:xh1_z}.

The subsequent discussion of this subsection is entirely restricted to
the largest bubble ($r_s=100 \, {\rm Mpc}$) for which the bubble's
comoving radius grows to $100 \, {\rm Mpc}$ by $\tau=10^8 \,
{\rm yr}$ when $\xh1$ is constant at $0.6$.  The bubble's apparent
image spans from $z=8.2$ to $7.6$, and  $\xh1$ changes  from
$0.64$ to $0.52$ across the bubble in the  evolving $\xh1$  model that
we have considered.  For the same model, $\xh1$ is $0.7$ at the
redshift where  the bubble is born if the quasar has an age of
$\tau_Q=1.3 \times 10^8 \, {\rm yr}$ at $z=8$.
\begin{figure}
\psfrag{r}[c][c][1][0]{{\LARGE $r(\tau)/r_s$ }}
\psfrag{tau}[c][c][1][0]{{\LARGE $\tau /10^7 \,{\rm yr}$ }}
\includegraphics[width=.8\textwidth, angle=-90]{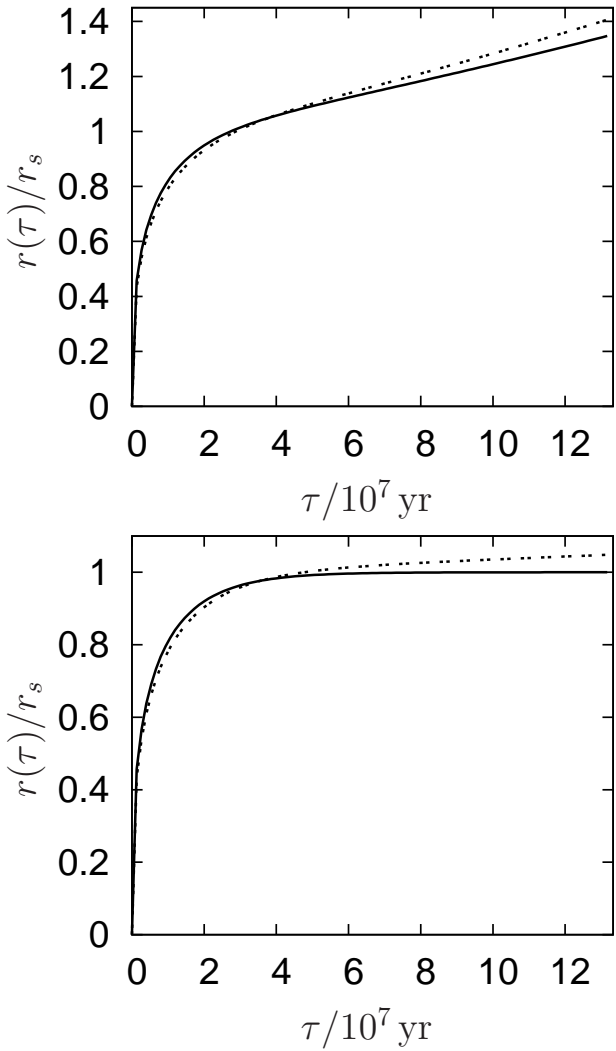}
\caption{This dotted curve shows the growth of an \HII bubble for 
  our model of evolving global neutral fraction. The 
  quasar age is set to $\tau_Q = 1.3 \times 10^8\,{\rm yr}$  at
  $z=8$.  The results for a constant neutral fraction $\xh1=0.6$ are
  shown for comparison (solid curve). The bottom and top panels
  respectively correspond to Models $(i)$ and $(ii)$.}
\label{fig:r_tau_com}
\end{figure}
\begin{figure}
\psfrag{eta}[c][c][1][0]{{\LARGE $\eta$ }}
\psfrag{s}[c][c][1][0]{{\LARGE $s$ }}
\psfrag{tauq}[c][c][1][0]{{\LARGE $\tau_Q /10^7 \,{\rm yr}$ }}
\psfrag{(i)}[c][c][1][0]{{\LARGE $(i)$ }}
\psfrag{(ii)}[c][c][1][0]{{\LARGE $(ii)$ }}
\includegraphics[width=.8\textwidth, angle=-90]{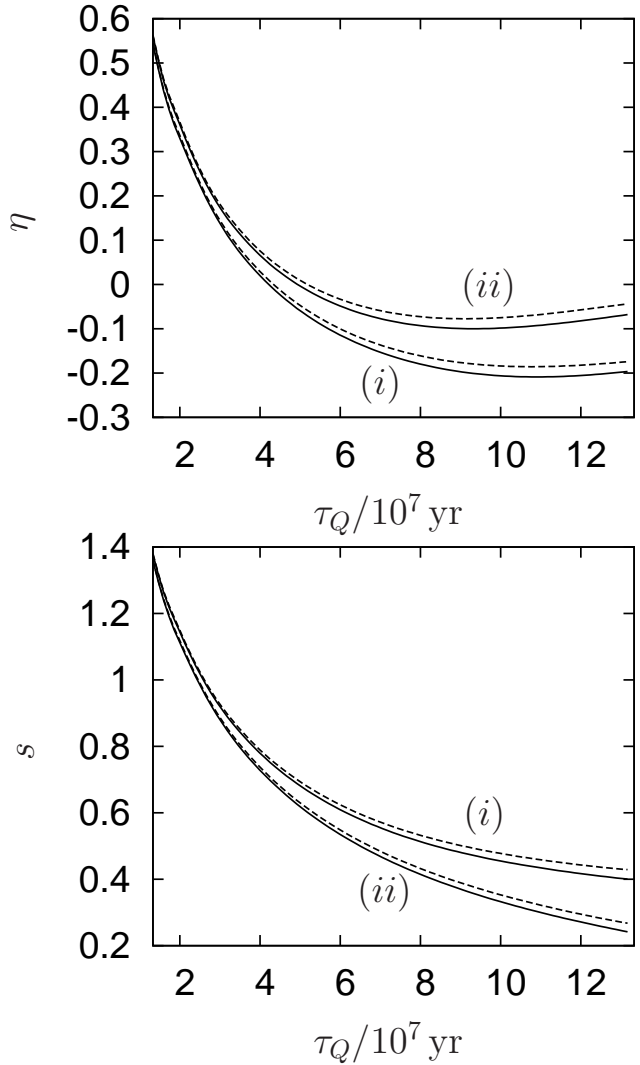}
\caption{This shows the variation of $\eta$ (top panel) and $s$
  (bottom panel)
  with $\tau_Q$ for an \HII bubble with $r_s=100$ Mpc. In both panels
the solid and dashed curves represent the constant  and evolving
  $\xh1$   models respectively. The two different sets of curves in
  each   panel   correspond to Models (i) and (ii) respectively.  }  
\label{fig:eta_s}
\end{figure} 

We have modified eq. (\ref{eq:yu}) to incorporate the evolution of
$\xh1$ (see Appendix \ref{ap:xh1} for details) for which  Figure
\ref{fig:r_tau_com} shows the growth of the bubble's comoving radius. 
For both Models $(i)$ and $(ii)$ we find that incorporating the
evolution of $\xh1$ initially slows the growth of the bubble relative
to the situation where $\xh1$ is constant. This is reversed in the
later stages where the growth is enhanced with respect to the
situation where $\xh1$ is constant. For the parameter range that we
have considered, the difference in  the bubble radius due to the
evolution of $\xh1$ is found to be less than $5 \%$.  Further, there
is a significant difference only at the later stages of the bubble's
growth $(\tau>3 \tau_{rec} \sim 4 \times 10^7 \, {\rm yr})$.  We next
consider the effect of the evolution of $\xh1$ on the anisotropy  and
shift parameters $\eta$ and $s$ (Figure \ref{fig:eta_s}).
We find that the effect is maximum
when the bubble is seen in the late stage of its growth where the
apparent image is compressed along the line of sight. The evolution of
$\xh1$ acts to reduce the compression and increase the shift relative
to a situation where $\xh1$ is constant. The maximum change is found
to be around $10 \%$.  This change in the anisotropy has a very small
impact on matched filter bubble detection. The radius of the best
matched filter, we find, changes by less than $5 \%$ whereas  the
corresponding SNR changes by less than $1 \%$.   The changes being
extremely small, we have not explicitly shown these results. 

The discussion till now has focused on a large bubble with $r_s=100\,
{\rm Mpc}$. The effect of the evolution of $\xh1$ will be less for
smaller bubbles.

\section{Summary  \& Conclusions}
The apparent image of a  quasar generated, spherical ionized region
appears anisotropic due to the finite
 light travel time  and the evolution of the global neutral fraction.   
We have quantified this  anisotropy and  assessed its impact on
matched filter bubble detection  in redshifted 21-cm maps at $z=8$. We
find that the anisotropy is less than  $20 \%$ for most of the
parameter space that we have considered. Further, the bubble appears
compressed along the LOS for nearly the entire parameter range where a
detection  is possible. In addition to the anisotropy,  the bubble's
center is also shifted from  the actual position of the quasar along
the LOS.  This shift, which  can  be rather large and may  exceed the
bubble radius. typically varies across  the range $20-80 \%$.
We find that these effects do not cause a significant  degradation of 
the SNR  in  matched filter bubble search  and  spherical filters
are adequate  for a preliminary  bubble detection.  We show that 
such a detection where the radius of the best match filter, its center
and the quasar redshift are all known can be used to estimate $\tau_Q$
the age of the quasar. This detection also provides an estimate of 
the neutral fraction $\xh1$  provided the quasar luminosity is known. 
 Follow up observations, subsequent to a detection, can in principle
 be used to measure the anisotropy which would impose further
 constrains on the quasar luminosity, $\tau_Q$ and $\xh1$.  
The global neutral fraction is not expected to evolve  very
rapidly with $z$ around  $z=8$, and this make a rather small change
$(< 10 \%)$ for the anisotropy and the shift relative a situation
where $\xh1$ is constant. 

An earlier work (Paper III) has shown that both the GMRT or the MWA are
most sensitive to the signal from an ionized bubble if it is at a
redshift $z \sim 8$.   At this $z$ a  bubble of  comoving radius  $\ge
24$ Mpc and $\ge 33$ Mpc  can  respectively be detected at $3\sigma$
and $5\sigma$ significance  with  $1000$ hrs of observation.  This,
however, presupposes the presence of an ionized bubble located near
the center of the FoV. It is, however, necessary to take into account
the abundance of ionized bubbles when considering possible
observational strategies.  

How many  \HII bubbles do we expect to  find above the detection
threshold in a single GMRT observational  volume at redshift $z=8$?
There is, till date, no quasar detected with $z>6.4$, though there are
galaxies which have been detected at $z>8$ (eg.  \cite{lehnert}) and
it is quite likely that quasars may  be detected in the near future. 
At present one thus typically uses  theoretical models which correctly
predict the 
observed  quasar abundance and  evolution  at $z\le 6$ to predict  
the expected number of  detectable quasar \HII bubbles at higher
redshifts, though these predictions are highly model dependent and
uncertain   \citep{rhook06}. Predictions  based on the CDM merger
driven model  \citep{wyithe03,wyithe05,rhook06} indicate that we
expect $\sim 1$ ionized bubble of comoving  radius  $\ge 20$ Mpc around an
active quasar in $400$ degree$^2$ and   $16 $ MHz volume.  This means
that the probability of  a detection is less than $5 \%$ in a  single
random GMRT observation volume  ($\sim 10 $ degree$^2$ and  $32 $ MHz)
at redshift $z=8$. Including the possibility of fossil bubbles whose
quasars have switched off increases the  number of \HII bubbles
\citep{wyithe05}, and we expect $\sim 5$ detectable bubbles  in a
random GMRT observation.  Note that, the shape of these  
fossil bubbles would not be distorted due to the FLTT  as  they have
ceased to grow.  

The typical size of quasar generated \HII bubbles at $z\sim8$ is
an important factor in matched filter bubble detection. 
\citet{maselli07} predict  the comoving radius of quasar generated
\HII regions to be $\sim 45$ Mpc at $z=6.1$ with $\xh1 = 0.1$.
This size is expected to be less at $z \sim 8$ where the  global
neutral fraction is higher. The typical  radius  of a galaxy generated
bubble is  expected  to be $\sim 10 $ comoving Mpc at $z \sim 8$
\citep{mcquinn07}.  These galaxy generated bubbles outside the 
quasar generated bubble that is targeted for detection will contribute
to the fluctuations in the matched filter estimator $\hat{E}$.
However, simulations (Paper II) indicate  that small bubbles of size
$10 \, {\rm Mpc}$ will not have much  impact on detection in the range
$(\ge 24 \, {\rm   Mpc})$ where a detection is feasible. 

Finally, we note that there are various other sources of anisotropy 
such as the spatial fluctuations in the neutral
hydrogen distribution due to galaxy generated  generated ionized
bubbles, anisotropic emission from the quasar itself which have been
ignored. galaxy generated generated ionized bubbles in the surrounding
IGM can cause significant anisotropies in the quasar generated ionized 
bubbles. One needs to use detailed simulation to investigate this
which is beyond the scope of this paper. We plan to 
address these issues in future.

\section{acknowledgments}
We would like to thank the anonymous referee for providing us with 
constructive comments and suggestions which helped to improve the paper. 
SM would like to thank Prasun Dutta, Prakash Sarkar, Tapomoy Guha
Sarkar, Subhasis Panda, Abhik Ghosh and Sanjit Das for useful discussions. 
KKD is grateful for financial support from Swedish Research Council (VR) 
through the Oscar Klein Centre.

\appendix

\section{Growth Equation for \HII Bubble Considering Evolving 
Neutral Fraction}
\label{ap:xh1}

Here we consider the growth of an \HII bubble embedded in a medium whose 
neutral fraction $\xh1$ outside the bubble is evolving.  The evolution of
 the neutral fraction is parameterized using 
\begin{equation}
\xh1(\tau)= \xh1(\tau_Q) \, f(\tau)
\label{eq:xh1}
\end{equation}
where $\xh1(\tau_Q)$ is the neutral fraction at the redshift $z_Q=8$,
and the function $f(\tau )$ which  quantifies the 
evolution satisfies $f(\tau_Q) =1$.  
Further, the ionizing photon emission rate of the quasar is expressed 
as $\dot{N}_{phs}=\dot{N}_{phs,i} \, s(\tau)$ with  $s(\tau)= 1 $ for 
Model $(i)$ and   $s(\tau)=\exp(\tau/\tau_s)$ for Model $(ii)$. 
Using these, the equation governing the growth of an ionized bubble
(eq. \ref{eq:yu}) may be written as 
\begin{equation}
 \frac{d}{d\tau} \left( r^3 \, f \right)=  \frac{r_s^3 s}{\tau_{rec}} 
- \frac{r^3}{\tau_{rec}}
\label{eq:grow_2}
\end{equation}
where $\tau_{rec}$ and $r_s$ are constants calculated 
using eq. (\ref{eq:taurec}) and  eq. (\ref{eq:rs}) respectively using 
$\xh1(\tau_Q)$ i.e. value of $\xh1$ at $z=8$.  
Note that we have
ignored the effect of the finite light travel time from the quasar to
the ionization front on $f(\tau)$. 
We then have the
solution 
\begin{equation}
r(\tau)= r_s \left\{\frac{1}{f(\tau)}\int\limits_0^{\tau} 
{\frac{d\tau^{\prime}}{\tau_{rec}} \exp \left[- \frac{1}{\tau_{rec}}
\int\limits_0^{\tau-\tau^{\prime}}  {\frac{d\tau^{\prime \prime}}
{f(\tau^{\prime \prime})}}\right]  s(\tau^{\prime})}\right\}^{\frac{1}{3}}
\label{eq:r_tau_mod} \,.
\end{equation}
The analysis  is considerably simplified if we approximate the evolution of
the neutral fraction as 
\begin{equation}
\frac{1}{f(\tau)} = 1 + b \, \left(\frac{\tau -
  \tau_Q}{\tau_{rec}}\right)  
\label{eq:f_tau}
\end{equation}
whereby eq. (\ref{eq:r_tau_mod})  can be written in terms of a single
integral as 
\begin{align}
r(\tau) =& r_s \left\{ \left[ 1 + b \, \left(\frac{\tau -
    \tau_Q}{\tau_{rec}}\right)\right] \, \int\limits_0^{\tau} \exp
\left[-\left(\frac{\tau - \tau^{\prime}}{\tau_{rec}}\right)
\right.\right.\nonumber\\
&\left.\left.\left(1 -  b\frac{\tau_Q}{\tau_{rec}}\right) -
\frac{b}{2}\left(\frac{\tau -  \tau^{\prime}}{\tau_{rec}}\right)^2
\right] s(\tau) \frac{d\tau^{\prime}}{\tau_{rec}}
\right\}^{\frac{1}{3}}
\label{eq:r_tau_f}
\end{align}
The integral in eq. (\ref{eq:r_tau_f}) can be easily
evaluated numerically. We use this equation to study the growth 
of an \HII bubble for a  wide range of values of the parameters $r_s$
and $\tau_Q\,$, taking into account the variation of the neutral
fraction. It is quite evident that we recover eqs. (\ref{eq:sol1}) and
(\ref{eq:sol2}) in the situation where $b=0$ {\it i.e.} the neutral
fraction does not evolve.  

The evolution of the neutral fraction $\xh1$ is largely unknown at the
redshifts $z \sim 8$.  For the purpose of our discussion we consider
two  models proposed by  \citet{choudhury05,   choudhury06}   which
implements most of the relevant physics  governing the thermal and
ionization history of the IGM through a   semi-analytical
formalism. Figure \ref{fig:xh1_z} shows the predicted evolution 
of $\xh1$ over the  relevant $z$ range.  For the  subsequent analysis
in this paper we have used  $b=0.015$ in eq. (\ref{eq:f_tau})  whereby 
the predicted $\xh1(z)$ is  in rough consistency with the evolution
allowed by the two models. 

The growth of an \HII bubble with evolving $\xh1$ (dotted line) has
been shown in Figure \ref{fig:r_tau_com}  for both Model $(i)$ and
$(ii)$ of photon emission rate.
For comparison with our previous analysis in this paper, growth for
constant $\xh1$ (solid line) is also shown in the same figure.


\begin{thebibliography}{99}
\bibitem[\protect\citeauthoryear{{Bharadwaj} \& 
{Pandey}}{2005}]{Bharadwaj05} 
 Bharadwaj, S., \& Pandey, S.~K.\ 2005, \mnras, 358, 968 

\bibitem[\protect\citeauthoryear{{Choudhury} \& 
{Ferrara}}{2005}]{choudhury05} 
Choudhury, T.~R., \& Ferrara, A., 2005, \mnras, 361, 577

\bibitem[\protect\citeauthoryear{{Choudhury} \&
  {Ferrara}}{2006}]{choudhury06}
{Choudhury}, T.~R.,  {Ferrara}, A., 2006, Cosmic Polarization, 
Editor - R. Fabbri(Research Signpost), p. 205, arXiv:astro-ph/0603149

\bibitem[\protect\citeauthoryear{Choudhury}{2009}]{choudhury09}
Choudhury, T.~R.,  2009, Current Science, 97, 6, 841


\bibitem[\protect\citeauthoryear{Datta, Bharadwaj \& Choudhury}{2007}]{datta2}
Datta, K.~K., Bharadwaj, S., \& Choudhury, T.~R., 2007,\mnras, 382, 109  

\bibitem[\protect\citeauthoryear{Datta, Majumdar, Bharadwaj \& Choudhury}
{2008}]{datta3}
Datta, K.~K., Majumdar, S., Bharadwaj, S., \& Choudhury, T.~R., 2008,\mnras,
 391, 1900  

\bibitem[\protect\citeauthoryear{Datta, Bharadwaj \& Choudhury}{2009}]{datta09} 
Datta, K.~K., Bharadwaj, S., \& Choudhury, T.~R.\ 2009, \mnras, 399, L132 

\bibitem[\protect\citeauthoryear{Furlanetto, Zaldarriaga \& Hernquist}
{2004}]{furlanetto1}
{Furlanetto}, S.~R., {Zaldarriaga}, M. \& {Hernquist}, L. 2004, \apj, 613, 1

\bibitem[\protect\citeauthoryear{Furlanetto, Oh \& Briggs}{2006}]{furlanetto2}
{Furlanetto}, S.~R., {Oh}, S.~P. \& {Briggs}, F.~H., Phys. Rep. 2006, 433, 181



\bibitem[\protect\citeauthoryear{Geil \& Wyithe}{2008}]{geil1}
 Geil, P.~M., \& Wyithe, J.~S.~B.\ 2008, \mnras, 386, 1683 

\bibitem[\protect\citeauthoryear{Geil et al.}{2008}]{geil2} 
Geil, P.~M., Wyithe, J.~S.~B., Petrovic, N., \& Oh, S.~P.\ 2008,
\mnras, 390, 1496  

\bibitem[\protect\citeauthoryear{Lehnert et al.}{2010}]{lehnert}
  Lehnert, M.~D., et al.\  2010, \nat, 467, 940

\bibitem[\protect\citeauthoryear{Maselli et al.}{2007}]{maselli07}
  Maselli, A., Gallerani,  S., Ferrara, A., \& Choudhury, T.~R.\ 2007,
  \mnras, 376, L34  

\bibitem[\protect\citeauthoryear{McQuinn et al.}{2007}]{mcquinn07} 
McQuinn, M., Lidz, A., Zahn, O., Dutta, S., Hernquist, L., \&
Zaldarriaga, M.\ 2007, \mnras, 377, 1043  


\bibitem[\protect\citeauthoryear{Rhook \& Haehnelt}{2006}]{rhook06}
Rhook, K.~J., \& Haehnelt, M.~G.\ 2006, \mnras, 373, 623 


\bibitem[\protect\citeauthoryear{Sethi \& Haiman}{2008}]{sethi08} Sethi, S., 
\& Haiman, Z.\ 2008, AJ, 673, 1S

\bibitem[\protect\citeauthoryear{Shapiro \& Giroux}{1987}]{shapiro87} 
Shapiro, P. R., \& Giroux, M. L.\ 1987, \apl, 321L, 107S

\bibitem[\protect\citeauthoryear{Shapiro et al.}{2006}]{shapiro06} 
Shapiro, P.~R., Iliev, I.~T., Alvarez, M.~A., \& Scannapieco, E.\ 2006,
 \apj, 648, 922

\bibitem [\protect\citeauthoryear{Swarup et al.}{1991}]{swarup}  
Swarup G., Ananthakrishnan S., Kapahi V.K., Rao A.P., Subramanya
C.R., Kulkarni V.K.,1991 Curr.Sci.,60,95

\bibitem[\protect\citeauthoryear{White, Becker, Fan \&  Strauss}{2003}]
{white03} White, R. L., Becker, R. H., Fan, X., \& Strauss, M. A.\ 2003,
 AJ, 126, 1

\bibitem[\protect\citeauthoryear{Wyithe \& Loeb}{2003}]{wyithe03}  
Wyithe, J.~S.~B., \& Loeb, A., 2003, ApJ, 595, 614 

\bibitem[\protect\citeauthoryear{Wyithe \& Loeb}{2004}]{wyithe04} 
Wyithe, J.~S.~B., \& Loeb, A.\ 2004, \apj, 610, 117

\bibitem[\protect\citeauthoryear{Wyithe, Loeb \& Barnes}{2005}]{wyithe05} 
Wyithe, J.~S.~B., Loeb, A., \& Barnes, D.~G.\ 2005, \apj, 634, 715 

\bibitem[\protect\citeauthoryear{Yu \& Tremaine}{2002}]{yu02}
 Yu, Q., \& Tremaine, S.\ 2002, \mnras, 335, 965 

\bibitem[\protect\citeauthoryear{Yu \& Lu}{2004}]{yu04}
 Yu, Q., \& Lu, Y.\ 2004, \apj, 610, 93 
 
\bibitem[\protect\citeauthoryear{Yu \& Lu}{2005}]{yu05b} 
Yu, Q. \& Lu, Y. \ 2005, \apj, 620, 31
 
\bibitem[\protect\citeauthoryear{Yu}{2005}]{yu05} 
Yu, Q. \ 2005, \apj, 623, 683

\end{thebibliography}
\end{document}